# Multiphysics-decision tree learning for improved variably saturated subsurface parameter estimation and reduced-order simulation

Michael J Friedel[1,2,3*] , Massimo Buscema[3,4]

[1] University of Colorado, Denver, Colorado, United States (michael.friedel@ucdenver.edu)

[2] *Research Corporation of the University Hawaii, University of Hawaii, Honolulu, HI, USA*

[3] *Semeion Research Institute, Via Sersale 117, 00128 Rome, Italy*

[4] *Mathematical and Statistical Sciences, University of Colorado, Denver, CO, USA*

**Abstract**

We present a multiphysics-decision tree learning algorithm for (1) estimating saturated hydraulic conductivity, thermal conductivity, bulk density and longitudinal dispersivity in the variably saturated subsurface governed by explicitly coupled equations for water, heat, and solute transport; and (2) providing reduced order simulation of time-dependent pressure head, temperature, and concentration with subsurface properties and/or changing surface boundary conditions. Because of the difficulty and costs associated with measuring subsurface water, heat, and solute transport properties in the field, this study promotes the use of available pressure head, temperature, and concentration measurements. We demonstrate that the proposed algorithm results in about one order of magnitude less error in estimated parameters than the traditional multiphysics numerical inversion. We further show that the multiphysics-decision tree learning algorithm reduces the computational burden associated of traditional parameter estimation with reductions in the number Jacobian sensitivity calculations by as much as 90% and the number of iterations required for convergence by up to an order of magnitude. A natural outcome following convergence of the proposed learning algorithm is the reduced order set of supervised decision tree learning models for predicting the pressure head (Random Forest), and the temperature and concentration (Ensemble Gradient Boosting) given knowledge of time, depth, and remaining pair of state variables. The supervised reduced-order modeling is extended to unsupervised machine learning for the simultaneous prediction of state variables by training a Self-Organized Map using the joint multiphysics-decision tree learning property estimates, stochastic boundary conditions, and subsurface state field measurements. The reduced-order machine learning models provide a computationally efficient alternative for studying the effects of changing subsurface water, heat, and solute transport properties and/or surface boundary conditions on coupled subsurface pressure head, temperature, and concentration.

Corresponding author: Email: address: mfriedel@hawaii.edu

[1] University of Colorado, Denver, Colorado, United States (michael.friedel@ucdenver.edu)

## 1. INTRODUCTION

Numerical models are widely used for simulating the transport of mass and energy in variably saturated subsurface systems (Langerbraber and Šimunek, 2005; Šimunek et al., 2012; Ebbrahimian et al., 2013). These models require the estimation of many disparate and spatially distributed parameters to capture field properties and processes that cannot otherwise be explicitly represented (Abbasi, et al., 2003, Finsterle et al., 2012). What makes subsurface parameter estimation so challenging is the ill-posed and nonunique inverse problem that arises due to a lack of information necessary to make the problem well defined (Elsheikh et al., 2012; Friedel and Iwashita, 2013; Petvipusit et al. 2014; Yeh et al., 2015). For example, typically there is a lack of physical properties available because of the difficulty and costs associated with measuring subsurface water, heat, and solute transport properties in the field. Instead, we acknowledge that more easily observed pressure head, temperature, and concentration measurements are available. Combining multiple state measurements (e.g. pressure head, temperature, concentration) in a multiphysics inverse framework is known to reduce the nonuniqueness and associated uncertainty in parameter estimates through the crossover of mutual information (Friedel, 2005; REF). The crossover of information shared among state variables (Learned-Miller, 2013) enhances Jacobian sensitivities and therefore the estimation of associated parameters when inverting the set of explicitly coupled governing equations (Eqn. 1-3). Evaluating the contributions of state variables to the sensitivity associated with a given parameter reveals the measure of dependency among random variables (i.e., mutual information (Friedel, 2005; REF). While the crossover effects of state variables benefit the variably saturated subsurface estimation process by enhancing parameter sensitivities, there remains other numerical modeling challenges (Carrera et al., 2005; Stuart, 2010), such as (1) identifying influential parameters, (2) identifying calibration measurements with mutual information about parameters, (3) knowledge or estimation of boundary conditions, (4) minimization of complex objective function topologies, and (5) computational resources required for field-scale multiphysics subsurface simulations.

Given the numerical challenges just mentioned, the field-scale modeling of subsurface water, heat, and solute transport becomes difficult and often impractical. For this reason, machine learning methods are sometimes used as an alternative modeling approach to solve three classes of hydrology problems: data-driven discoveries, data-driven solutions, and combined data-driven discovery and data-driven solutions. Examples of data-driven discoveries include evaluating mutual information content, such as nonlinear relations and among parameters (state variables and properties) using cross-component plots (Vesanto and Alhoniemi, 2000) of network weights arrived at using a competitive learning algorithm (Kohonen, 2001); and evolving predictive equations from field data using a genetic programming approach, such as symbolic regression (Babovic and Keijzer, 2000). Examples of data-driven solutions include using *supervised* machine learning to predict groundwater recharge (Huang, et al., 2019), predict groundwater nitrate concentration (Ransom, et al., 2017), and classify nitrate contamination (Juntakut, et al., 2019), and predict groundwater arsenic concentrations (Tan et al., 2020); and *unsupervised* machine learning to estimate continuous hydrogeophysical properties (Friedel, 2016). In a recent study, the efficacy of supervised and unsupervised machine learning methods was evaluated for their ability to classify and predict groundwater nitrate status (Friedel et al., 2020). Learn heuristics (metaheuristics in supervised machine learning) are sometimes used for the identification of influential parameters (Buscema, 2004), such as in groundwater-quality models (Friedel and Buscema, 2016; REF). Other deep learning algorithms have been used to forecast groundwater levels (Bowes et al., 2019, Afzall et al., 2020; REF) and predict groundwater recharge (Huang, et al., 2019). Currently, there is growing consensus that machine learning methods relying on data need to respect the laws of physics.

Physics-informed machine learning algorithms are trained to solve learning tasks while respecting the laws of physics (Yang et al., 2019, Jagtap et al., 2020, Kadeethum et al., 2020). Most studies involve the application of deep neural networks to learn solutions and parameters in differential equations (Lagaris et al., 1998; Weinan and Yu, 2018; Raissi et al, 2019, Jagtap et al., 2020). Recently, physics-informed deep neural networks are extended to steady-state inverse problems associated with partial differential equations for estimating hydraulic conductivity using sparse sets of measurements, such as hydraulic conductivity and hydraulic head (Tartakovsky, et al., 2020) and hydraulic conductivity, hydraulic head, and concentration (He et al., 2020). Deep neural networks also are combined with the finite-element method for estimating parameters in partial differential equations with state variables that were fully known (Xu and Darve, 2020). These studies demonstrate that physics-informed deep learning networks can improve the accuracy of parameter estimates as compared to deep neural networks trained only with data.

In this study, the *aim* is to improve the simulation process for enhanced understanding of coupled transport of water-heat-solutes in the variably saturated zone. Our objectives are twofold. First, improve the variably saturated subsurface parameter estimation process by enhancing speed and reducing uncertainty; second, develop a reduced order model for rapid and computationally efficient assessment of subsurface transport properties and/or changing surface boundary conditions on subsurface state variables. We *hypothesize* that it is possible to identify and apply influential parameters for use in: (1) the multiphysics-decision tree learning inversion for simultaneous estimation of water, heat, and solute transport properties, and (2) the development of a reduced-order model based on training of the unsupervised learning model with these same parameters. To test this hypothesis, we develop and apply a joint multiphysics-decision tree learning algorithm: multiphysics numerical model and surrogate machine learning model using field measurements and multicomponent objective function. The field data reflects a stochastic set of surface and subsurface measurements. The multiphysics numerical model component comprises an explicitly coupled set of governing partial differential equations for water, heat, and solute transport that maximize mutual information in the state variables (Friedel, 2005). The surrogate machine learning component comprises one (or more) open source, well documented, and well referenced *supervised* machine learning (e.g., Random Forest, Gradient Boosting, Ensemble Gradient Boosting, and Gaussian Process) and *unsupervised* machine learning (e.g. Self-Organizing Map) models. The *objectives* are to quantify and compare known and estimated multiphysics (water, heat, and solute) transport properties and to compute reduced-order time-dependent simulations of state variables (pressure head, temperature, concentration) in the heterogeneous (layered) variably saturated subsurface. The estimated multiphysics transport properties and reduced order time-dependent state variable simulations are arrived at following calibration of the multiphysics-decision tree learning model.

## 2. MATERIALS AND METHODS

We present a multiphysics-decision tree learning algorithm that combines stochastic field measurements, multiphysics numerical model, surrogate machine learning models, and multicomponent objective function. This algorithm uses nonlinear regression to minimize differences in field measured and model simulated state variables (pressure head, temperature, and concentration) arrived at using the multiphysics numerical and surrogate machine learning models (Figure 1). The three primary components of this algorithm are briefly described next.

### 2.1 Multiphysics numerical model

The multiphysics numerical model used in the algorithm comprises a set of explicitly coupled partial differential equations that describe variably saturated water-heat-solute transport (after Friedel, 2000, 2005). The explicit coupling of these partial differential equations enhances parameter sensitivities by maximizing mutual information in state variables thereby reducing uncertainty in the estimation of transport properties. The governing equation for two-dimensional *water transport* subject to pressure head, temperature, and concentration gradients in variably saturated geologic media is given by

$$\rho_L \Lambda \frac{\partial \theta_L}{\partial t} + \rho_L \frac{\partial \theta_V}{\partial t} = \frac{\partial}{\partial x}\left(D_{\theta V}\rho_L C_\psi \frac{\partial \psi}{\partial x}\right) + \frac{\partial}{\partial x}\left(\rho_L K_{xx} \frac{\partial \psi}{\partial x}\right) + \frac{\partial}{\partial x}\left(\rho_L K_{xy} \frac{\partial \psi}{\partial y}\right) + \frac{\partial(\rho_L K_{xy})}{\partial x} + \frac{\partial}{\partial x}\left(D_{TV}\rho_L \frac{\partial T}{\partial x}\right) +$$
$$\frac{\partial}{\partial x}\left(D_{TLxx}\rho_L \frac{\partial T}{\partial x}\right) + \frac{\partial}{\partial x}\left(D_{TLxy}\rho_L \frac{\partial T}{\partial y}\right) - \frac{\partial}{\partial x}\left(D_{CV}\rho_L \frac{\partial C}{\partial x}\right) - \frac{\partial}{\partial x}\left(D_{CLxx}\rho_L \frac{\partial C}{\partial x}\right) - \frac{\partial}{\partial x}\left(D_{CLxy}\rho_L \frac{\partial C}{\partial y}\right) +$$
$$\frac{\partial}{\partial y}\left(D_{CV}\rho_L C_\psi \frac{\partial \psi}{\partial y}\right) + \frac{\partial}{\partial y}\left(\rho_L K_{yx} \frac{\partial \psi}{\partial x}\right) + \frac{\partial}{\partial y}\left(\rho_L K_{yy} \frac{\partial \psi}{\partial y}\right) + \frac{\partial(\rho_L K_{yy})}{\partial y} + \frac{\partial}{\partial y}\left(\rho_L D_{TV} \frac{\partial T}{\partial y}\right) + \frac{\partial}{\partial y}\left(\rho_L D_{TLyx} \frac{\partial T}{\partial x}\right) +$$
$$\frac{\partial}{\partial y}\left(\rho_L D_{TLyy} \frac{\partial T}{\partial y}\right) - \frac{\partial}{\partial y}\left(\rho_L D_{CV} \frac{\partial C}{\partial y}\right) - \frac{\partial}{\partial y}\left(\rho_L D_{CLyx} \frac{\partial C}{\partial x}\right) - \frac{\partial}{\partial y}\left(\rho_L D_{CLyy} \frac{\partial C}{\partial y}\right) + \rho_L Q_{Lp} + \rho_L Q_{Ld} \quad (1)$$

where Ψ [m], T [$^0$C], and C [mol-kg$^{-1}$] are state variables, C = solute concentration [mol kg$^{-1}$], C$_\psi$ = moisture capacity [m$^{-1}$], D$_{\theta V}$ = isothermal vapor diffusivity [m$^2$ d$^{-1}$], D$_{TL}$ = thermal liquid diffusivity [m$^2$ d$^{-1}$ $^0$C$^{-1}$], D$_{TV}$(Ψ, T, C) = vapor thermal diffusivity [m$^2$ d$^{-1}$ $^0$C$^{-1}$], D$_{CV}$ = vapor diffusivity in soil because of solute concentration [m$^2$ d$^{-1}$ kg mol$^{-1}$], D$_{CL}$ = liquid water diffusivity in soil because of solute concentration [m$^2$ d$^{-1}$ kg mol$^{-1}$], , K$_{xx}$ = principal component of hydraulic conductivity in xx direction [m d$^{-1}$], K$_{xy}$ = principal component of hydraulic conductivity in xy direction [m d$^{-1}$], K$_{yy}$ = principal component of hydraulic conductivity in yy direction [m d$^{-1}$], Q$_{Lp}$, Q$_{Ld}$ = liquid water flux point and distributed source or sink [m d$^{-1}$]; t = time [d], T = temperature [$^0$C], x,y = cartesian coordinates (horizontal, vertical), Λ(Ψ, T, C) = dimensionless storativity-type term, θ$_L$ = liquid water (moisture) content [cm$^3$ cm$^{-3}$], θ$_V$ = water vapor content [cm$^3$ cm$^{-3}$], $\rho_L$ (Ψ, T, C) = density of liquid [kg m$^{-3}$], and Ψ = pressure head [m].

The governing equation for two-dimensional *heat transport* subject to pressure head, temperature, and concentration gradients in variably saturated geologic media is given by

$$f_1 \frac{\partial T}{\partial t} + f_2 \frac{\partial \theta_L}{\partial t} + f_3 \frac{\partial C}{\partial t} = \frac{\partial}{\partial x}\left(\sigma' \frac{\partial \psi}{\partial x}\right) + \frac{\partial}{\partial x}\left(C_L T K_{xx} \frac{\partial \psi}{\partial x}\right) + \frac{\partial}{\partial x}\left(C_L T K_{xy} \frac{\partial \psi}{\partial y}\right) + \frac{\partial}{\partial x}\left(\sigma \frac{\partial T}{\partial y}\right) +$$
$$\frac{\partial}{\partial x}\left(C_L T D_{TLxx} \frac{\partial T}{\partial x}\right) + \frac{\partial}{\partial x}\left(D_{TLxy}\rho_L \frac{\partial T}{\partial y}\right) - \frac{\partial}{\partial x}\left(\sigma'' \frac{\partial C}{\partial x}\right) - \frac{\partial}{\partial x}\left(C_L T D_{CLxx} \frac{\partial C}{\partial x}\right) - \frac{\partial}{\partial x}\left(C_L T D_{CLxy} \frac{\partial C}{\partial y}\right) +$$
$$\frac{\partial}{\partial y}\left(\sigma' \frac{\partial \psi}{\partial y}\right) + \frac{\partial}{\partial y}\left(C_L T K_{yx} \frac{\partial \psi}{\partial x}\right) + \frac{\partial}{\partial y}\left(C_L T K_{yy} \frac{\partial \psi}{\partial y}\right) + \frac{\partial}{\partial y}\left(\sigma \frac{\partial T}{\partial y}\right) + \frac{\partial}{\partial y}\left(C_L T D_{TLyx} \frac{\partial T}{\partial x}\right) +$$
$$+ \frac{\partial}{\partial y}\left(C_L T D_{TLyy} \frac{\partial T}{\partial y}\right) - \frac{\partial}{\partial y}\left(\sigma'' \frac{\partial C}{\partial y}\right) - \frac{\partial}{\partial y}\left(C_L T D_{CLyx} \frac{\partial C}{\partial x}\right) - \frac{\partial}{\partial y}\left(C_L T D_{CLyy} \frac{\partial C}{\partial y}\right) + C_L T \frac{\partial K_{xy}}{\partial x} + C_L K_{xy} \frac{\partial T}{\partial x} +$$
$$C_L T \frac{\partial K_{yy}}{\partial y} + C_L K_{yy} \frac{\partial T}{\partial y} + Q_{Lp} + Q_{Ld} \quad (2)$$

where Cv = the global heat capacity of the porous medium [cal m$^{-3}$ $^0$C$^{-1}$], C$_L$ = volumetric heat capacity of liquid water [cal m$^{-3}$ $^0$C$^{-1}$], h$_r$ = the relative humidity [dimensionless]; $f_1(\Psi, T, C)$ = bulk heat capacitance type term [cal m$^{-3}$ $^0$C$^{-1}$], $f_2(\Psi, T, C)$ = bulk latent heat term [cal m$^{-3}$], $f_3(\Psi, T, C)$ = bulk latent heat type term [cal m$^{-3}$ kg mol$^{-1}$], L = the latent heat of vaporization [cal gm$^{-1}$] , $\sigma'$ = effective latent heat term associated with a pressure gradient [cal m$^{-3}$ $^0$C$^{-1}$], $\sigma''$ = effective latent heat term associated with a concentration gradient [cal m$^{-3}$ kg mol$^{-1}$], and ρ$_0$ = the saturated water vapor density [kg m$^{-3}$].

The governing equation for two-dimensional *solute transport* subject to pressure head, temperature, and concentration gradients in variably saturated geologic media is given by

$$C \frac{\partial \theta_L}{\partial t} + R \frac{\partial C}{\partial t} = \frac{\partial}{\partial x}\left(D_{CCxx} \frac{\partial C}{\partial x}\right) + \frac{\partial}{\partial y}\left(D_{CCyy} \frac{\partial \psi}{\partial y}\right) + \frac{\partial}{\partial x}\left(D_{CLxx} \frac{\partial C}{\partial y}\right) + \frac{\partial}{\partial y}\left(D_{CCyx} \frac{\partial C}{\partial x}\right) - \frac{\partial}{\partial x}\left(D_{C\psi xx}\rho_L \frac{\partial \psi}{\partial x}\right) -$$
$$\frac{\partial}{\partial y}\left(D_{C\psi yy}\rho_L \frac{\partial \psi}{\partial y}\right) - \frac{\partial}{\partial x}\left(D_{C\psi xy}\rho_L \frac{\partial \psi}{\partial y}\right) - \frac{\partial}{\partial y}\left(D_{C\psi yx}\rho_L \frac{\partial \psi}{\partial x}\right) + \frac{\partial}{\partial x}\left(D_{CT} \frac{\partial T}{\partial x}\right) + \frac{\partial}{\partial y}\left(D_{CT} \frac{\partial T}{\partial y}\right) - q_{Lx} \frac{\partial C}{\partial x} -$$
$$q_{Ly} \frac{\partial C}{\partial y} - V_{Lx} C C_\psi \frac{\partial \psi}{\partial x} - V_{Ly} C C_\psi \frac{\partial \psi}{\partial y} - \theta_L C \frac{\partial V_{Lx}}{\partial x} - \theta_L C \frac{\partial V_{Ly}}{\partial y} - \lambda C R + Q_{Cp} + Q_{Cd} \quad (3)$$

where D$_{CCxx}$ = moisture diffusivity because of solute concentration in xx direction [m$^2$ d$^{-1}$ kg mol$^{-1}$]; D$_{CCyy}$ = moisture diffusivity because of solute concentration in yx direction [m$^2$ d$^{-1}$ kg mol$^{-1}$], D$_{CCyy}$ = moisture diffusivity because of solute concentration in yx direction [$m^2$ $d^{-1}$ kg $mol^{-1}$], $D_{CT}$ ($\Psi$, T, C) = the solute diffusivity due to temperature gradient [$m^2$ $d^{-1}$ $^0C^{-1}$], $D_{C\psi yy}$, = diffusivity because of a pressure head gradient in yy direction [$m^2$ $d^{-1}$ $^0C^{-1}$], $D_{C\psi xy}$ = diffusivity because of a pressure head gradient in xy direction [$m^2$ $d^{-1}$ $^0C^{-1}$], $D_{C\psi xx}$ = diffusivity because of a pressure head gradient in xx direction [$m^2$ $d^{-1}$ $^0$ $C^{-1}$], $D_{CT} = D_{TS}\varepsilon\theta_L$ [$m^2$ $d^{-1}$ mol $kg^{-1}$ $^0C^{-1}$], R = the retardation factor [dimensionless], $V_{Lx}$, $V_{Ly}$ = average bulk solution velocities in the x- and y- directions [m $d^{-1}$], ε = the tortuosity factor [dimensionless], and λ = solute decay constant [$d^{-1}$].

These governing variably saturated water, heat, and solute transport (1–3) equations are solved numerically for a given set of initial and boundary conditions using the Galerkin finite-element approach (Friedel, 2000). Two characteristics that distinguish the finite-element method from other numerical procedures are the use of an integral formulation to generate a system of algebraic equations and continuous piecewise smooth functions for approximating the unknown quantities (state variables). The finite-element method comprises the following five basic steps. (1) The region is discretized into elements. This discretization includes locating and numbering the node points and coordinate values. (2) An approximation is specified, and equations written in terms of the unknown nodal values. (3) A system of nonlinear algebraic equations is developed. In using the Galerkin method, the weighting function for each unknown nodal value is defined and weighted residual integral evaluated. Application of this method generates one equation per state variable at each nodal value. Given that there are three state variables (pressure head, temperature, and concentration), three approximation equations are written for each node. (4) The system of approximating algebraic equations is then iteratively solved. These equations are intrinsically nonlinear because many parameters are unknown quantities that depend on the magnitude of one or more dependent variables. For this reason, the equations are quasi-linearized using a Picard approach prior to solution. (5) The water, heat, and solute transport properties are estimated using nonlinear least-squares regression. The reader is referred to Friedel (2000) for the complete derivation of approximating equations used herein.

**2.2 Surrogate machine learning models**

Surrogate models are considered computationally efficient models that approximate key characteristics of the full-order model governed by partial differential equations (Frangos et al. 2010). Surrogate modeling techniques are classified into three types: physics-based models (Durlofsky and Chen 2012; Josset et al. 2015), projection-based models (Fang et al. 2013; Lassila et al. 2014), and machine-learning based models (Frangos et al. 2010; Li et al. 2017). Physics-based surrogate models are derived from high-fidelity models using approaches by simplifying physics assumptions, coarsening of grids, and upscaling of model parameters (Frangos et al. 2010; He 2013; Xiao and Tian, 2020). Projection based reduced-order models are obtained by an orthogonal projection of the governing equations onto principal orthogonal decomposition modes. These two approaches result in a truncated numerical system representing the most salient characteristics of the governing equations (Amsallem and Farhat, 2012; Taira et al., 2017). These model reduction techniques have limitations especially for complex systems (Peherstorfer and Willcox, 2016). Machine learning offers an alternative to generate accurate parametric reduced-order models (Omer et al., 2019). Machine learning models will be developed using simulation data to regress relations between the input (predictor variables) and corresponding output (dependent variables) of interest (Wood 2018). In developing a surrogate machine learning model, the following five components are considered: (1) field measurements, (2) model selection, (3) feature selection, and (4) training and testing.

2.2.1 Field measurements

In this study, the field measurements are synthetic being generated based on a conceptual subsurface numerical model, application of stochastic boundary and initial conditions, and simulated sampling. The subsurface model (Figure 2) is conceptualized to reflect a managed aquifer recharge basin operated along the southern California Coastal Range (Munevar and Marino, 1999). In this model, water, heat, and solutes are assumed to percolate through the variably saturated and heterogeneous subsurface basin. The total depth from the streambed surface to the water table is 14 m and includes three anisotropic layers (top to bottom): highly conductive sand and gravel deposit (3-m-thick); less conductive, fine-grained silt deposit (2-m-thick); and fine-grained sand deposit with intermediate conductivity (9-m-thick). The assignment of water, heat, and solute transport properties in the managed aquifer recharge basin reflects findings from previous studies (Izbicki et al., 2000) that are summarized in Table 1. The respective hydraulic conductivity anisotropy ratios (vertical/horizontal) for the upper (gravel), middle (silt), and lower (sand) layers are 2, 0.9, and 1.5 indicating a preference for transport vertically in the upper and lower units and horizontal transport in the middle layer. The computational mesh used to represent the geologic framework is uniformly spaced at 1-m intervals in the vertical (15 nodes) and horizontal (3 nodes) directions.

Application of stochastic boundary conditions to the numerical model reflects an initially dry streambed surface that receives treated wastewater of known stage (or pressure head), temperature, and solute concentration. For this case, the surface-water characteristics reflect a normally distributed pressure head (from -5.2 m to 0.97 m), temperature (from 23 $^0$C to 39 $^0$C) and convective solute flux (Cauchy-type; Friedel, 2000) where the stream concentration (from 0.3 mol kg$^{-1}$ to 1.9 mol kg$^{-1}$) is considered the ambient concentration. Similar conditions are applied to the water table surface reflecting constant groundwater character. In this instance, a constant pressure head (0 m) and ground-water temperature (10 $^0$C) are applied along the lower boundary of the numerical model together with convective solute flux condition where the ground-water concentration (0.001 mol/kg) is considered the ambient background concentration. Cauchy-type conditions (water, heat, and solute) are applied along the vertical side boundaries with similar ambient conditions used, and the initial conditions (pressure head = -7.0 m, temperature =15 $^0$C, concentration =0.01 mol kg$^{-1}$) are applied uniformly with depth. The combination of stochastic upper boundary conditions and deterministic lower boundary conditions together with the initial conditions result in the simulation of subsurface pressure head, temperature, and concentration values at 26 locations over a period of about 3.8 years (Table 3).

Simulated sampling is undertaken assuming profiles reflecting nests of tensiometers (pressure head), thermocouples (temperature), and suction lysimeters (pore-fluid chemistry) located at 15 depths spaced 1-meter apart from the streambed surface to groundwater table. Sampling at these locations is assumed to occur over a 2-week period at the following times (in days): 0.01, 0.02, 0.03, 0.04, 0.05, 0.06, 0.07, 0.08, 0.09, 0.1, 0.2, 0.3, 0.4, 0.5, 0.6, 0.7, 0.8, 0.9, 1, 2, 3, 4, 5, 6, 7, 8, 9, 10, 11, 12, 13 and 14. The current sampling schedule captures early, middle, and late stage infiltration of water, heat, and solute from the dry surface to saturated water table. Maintaining the same measurement sampling schedule as in Friedel (2005) affords a direct comparison of estimated water, heat, and solute transport properties among approaches. By simulating 27 noise-free measurements at 15 locations, a total of 405 measurements per state variable (pressure head, temperature, and concentration) are available for the analysis and comparison (total of 1215 measurements). Correlated random variables (Myers and Yeh, 1999) are used to generate 1000 sets of stochastic boundary conditions (e.g., 42,000 normally distributed pressure head, temperature, concentration measurements over about 38 years) that are randomly down selected to 100 sets (e.g., 4,074 normally distributed pressure head, temperature, concentration values over about 3.8 years) for application to the numerical model. The statistical summaries of simulated state variables reveal good correspondence between original and down selected set of boundary conditions (Table 2).

2.2.2 Model selection

The purpose of model selection is to identify an appropriate machine learning algorithm for predicting one or more dependent variables as a function of independent predictor variables. The solution to regression problems using machine learning algorithms typically focus on predicting a single dependent response variable as a function of multiple independent predictor variables (also called features). In this study, the supervised machine learning algorithms (Pedregosa et al., 2011), e.g. Boosted Regression Trees (De'ath, 2007), Ensemble Gradient Boosting Regression (Boehmke et al., 2019), Gaussian Process Regression (Rasmussen and Williams, 2006), Gradient Boosting Regression (Friedman, 2001), and Random Forest Regression (Breiman, 2001), are evaluated to predict a single state variable as a function of time, distance, and two other state variables (e.g. pressure head as a function of temperature and concentration, temperature as a function of pressure head and concentration, and concentration as a function of pressure head and temperature). Likewise, an unsupervised machine learning algorithm, called Self-Organizing Map (Kohonen, 2001), is evaluated for the ability to predict multiple state variables (e.g. pressure head, temperature, and concentration) as a function of various subsurface parameters (location, time, and water, heat, and solute transport properties). For a review of the mathematics associated with these algorithms, the reader is referred to their associated references. The actual properties used in development of the unsupervised machine learning model is evaluated using the feature selection algorithm discussed next.

2.2.3 Feature selection

An important step prior to presenting input data to the unsupervised machine learning algorithm involves feature selection (Brown et al., 2012). The process of feature selection typically involves reducing the initial set of independent (predictor) variables to a set of influential features characterized by their mutual information (Gao et al., 2017) content using filter, wrapper, or embedded methods (Chandrashekar and Sahin, 2014). The resulting set of influential predictor variables are then used as input to train the learning algorithm (Gao et al., 2017).

In this study, the influential model features are identified based on their mutual information content using a learn heuristics algorithm. The learn heuristics algorithm has two components: Training With Input Selection and Testing TWIST (Buscema, et al., 2013). The input selection component of the algorithm operates as an evolutionary wrapper that reduces the dimensionality of data by extracting the minimum number of predictor variables necessary to maximize conservation of mutual information. The wrapper requires that the selection of best feature subset takes place while considering as relevant those attributes that allow the induction algorithm to generate a more accurate performance measure constrained by one or more dependent variables. The combination of input (feature) selection with training and testing requires an algorithm that can identify the best distribution of global data divided in two optimally balanced subsets containing a minimum number of input features useful for pattern estimation. The feature selection flowchart, shown in Figure 3, follows the following three steps (Buscema, et al., 2013):

    1. Initialize genetic population
    2. Perform an evolutionary loop
        a. Fitness evaluation of the proposal after splitting of individuals of the genetic population at generation (n) (From step a to step e)
        b. Crossover and offspring are produced
        c. Random mutation is applied

d. Setup of the new population

e. If average fitness continues to grow then restart from beginning; else terminate.

Each time an individual of the genetic population presents its hypothesis of splitting the global data set into two subsets A and B.

3. Save subset A and subset B with the best fitness.

From a practical view, the TWIST algorithm inserts supervised learning into an evolutionary algorithm. In this case, a series of Multilayer Perceptrons (MPL) with Back Propagation are proposed and subsequently evaluated using a genetic algorithm and fitness function constrained by the three state variables: pressure head, temperature, and solute concentration. The fitness function defines the relative importance of a design (e.g., the proposed set of MPL). The fitness function may be defined in several different ways, such as follows: $F_i = (1 + \text{epsilon}) f_{max} – f_i$, where the cost function (penalty function value for a constrained problem) for the $i^{th}$ design, $f_{max}$ is the largest recorded cost (penalty) function value, and epsilon is a suitably small value, e.g., $1 \times 10^{-6}$, to prevent numerical difficulties when $F_i$ becomes 0. In this study, there are 4 layers: 1 input, 3 hidden (determined by adding layers until the test error is stable; Bengio, 2012) and 1 output layer. The number of perceptrons in the input and hidden layers reflect the number of water, heat, and solute transport properties (features), whereas the number of output perceptrons is equal to the number of state variables (hypothesis). Suitable parameters used in the genetic algorithm were arrived at iteratively, e.g. number of chromosomes (binary encoding) = 100, crossover probability = 0.8, and mutation probability = 0.1.

Given that the genetic algorithm finds optimal solutions in the vicinity of the true global solution, the converged set of optimal features are nonunique representing one set of many possible combinations that may include local minima. For this reason, the algorithm is randomly restarted, and converged solutions recorded. Following the accumulation of converged models, the mode of occurrence is computed for each of the features. This process is repeated until the feature of occurrence for feature modes satisfy a user specified threshold (probability that the feature mode is important). For example, the user specified threshold of 0.67 (reflecting feature modes are within 1 standard deviation of the mean) is used to identified six robust features containing mutual information following several TWIST restarts (N=15): time, distance, hydraulic conductivity, thermal conductivity, bulk density, and dispersivity (Table 4). For a review of the mathematics associated with this algorithm, the reader is referred to the associated reference (Buscema, et al., 2013).

2.2.5 Training and testing

In building a surrogate machine learning model, the set of down selected stochastic field measurements (N=39,285 records) are randomly shuffled and split into sets for training (N=31,428, 80% of the field records) and independent testing (N=7,857 records, 20% split of the field records; N = 405 records, 1% split of the field records; and N = 80 records, 0.2% split of the field records) when using machine learning algorithms. The *supervised* machine learning algorithms (Pedregosa *et al.*, 2011) considered include Boosted Regression Trees (De'ath, 2007), Ensemble Gradient Boosting (Boehmke et al., 2019), Gaussian Processes (Rasmussen and Williams, 2006), Gradient Boosting (Friedman, 2001), and Random Forest (Breiman, 2001), . The feature importance for each of the trained pressure head, temperature, and solute concentration models reveal their sensitivity to time and distance and insensitivity to the four physical properties. For example, the feature importance plots for each of the Random Forest models reveal sensitivity to time (t) and depth (Y) but not the physical properties (Figure 4). Given the insensitivity of the supervised machine learning models to physical properties, these models are retrained to predict a single dependent response as a function of time, distance, and other two dependent responses, e.g., pressure head as a function of time, distance, temperature, and concentration; temperature as a function of time, distance, pressure head and concentration; and concentration as a function of time, distance, pressure head, and temperature. The corresponding feature importance plots reveal two important but different predictor variables for each of the Random Forest models, e.g., pressure head predictions are sensitive primarily to temperature and concentration, temperature predictions are sensitive primarily to concentration and distance, and concentration predictions are sensitive primarily to distance and time (Table 5).

The training and testing results of the supervised Random Forest machine learning model is presented in Figure 5. The four panels in each quadrant indicate testing on the full training set, and various random splits. The upper right quadrant reveals unbiased independent predictions of pressure head as a function of time, distance, temperature, and concentration. Similar unbiased prediction results are shown across the splits when predicting temperature as a function of time, distance, pressure head, and concentration (lower left) and when predicting concentration as a function of time, distance, pressure head, and temperature (lower right). By contrast, the Gaussian Process and Gradient Boosting resulted in biased predictions at extremes values of the selected state variables, e.g., the Gaussian Process produced biased predictions at high concentration values, whereas Gradient boosting produced biased predictions at low values of pressure head and concentration. The final criteria used in selecting a supervised algorithm for inclusion in the surrogate machine learning model component included achieving: a comparatively low number of estimators (hyperparameter tuning), low mean squared error, reasonable processing speed, and unbiased predictions. The learning algorithms satisfying these preferred criteria included: Random Forest for predicting pressure head optimal number of estimators: 24, root mean squared error: 0 0.26 m); and Ensemble Gradient Boosting for predicting temperature (optimal number of estimators: 885, root mean squared error: 1.1 $^0$C) and concentration (optimal estimators: 464, root mean squared error 0.03 mol kg$^{-1}$). The workflow used to identify suitable supervised surrogate models for predicting pressure head, temperature, and concentration as part of the joint multiphysics-decision tree learning algorithm is presented in Figure 6.

## 2.3 Multicomponent objective function

Solution to the joint multiphysics-decision tree inverse problem is achieved by applying nonlinear least-squares regression to minimize the associated multicomponent objective function (Figure 7). The multicomponent objective function includes a multiphysics numerical model component, a surrogate machine learning component, and Tikhonov regularization (Tikhonov and Arsenin, 1977) component given by

$$\min \varphi(\psi_m, T_m, C_m, \psi_{MP}, T_{Mp}, C_{MP}, \psi_{ML}, T_{ML}, C_{ML}, \ldots \psi_{ML}, T_{ML}, C_{ML}, d) = min[$$

$$\lambda \left\{ \sum_{i=1}^{NP} w_{MP\psi i} [\psi_{mi}(t,x,y) - \psi_{MPsi}(p)]^2 + \sum_{j=1}^{NT} w_{MPTj} [T_{mj}(t,x,y) - T_{MPsj}(p)]^2 + \sum_{k=1}^{NC} w_{MPCk} [C_{mk}(t,x,y) - C_{MPsk}(p)]^2 \right\}$$

$$+ \lambda \left\{ \sum_{i=1}^{NP} w_{ML\psi i} [\psi_{mi}(t,x,y) - \psi_{MLsi}(p)]^2 + \sum_{j=1}^{NT} w_{MLTj} [T_{mj}(t,x,y) - T_{MLsj}(p)]^2 + \sum_{k=1}^{NC} w_{MLCk} [C_{mk}(t,x,y) - C_{MLsk}(p)]^2 \right\}$$

$$+ \sum_{l=1}^{NR} w_{rl} [d_{ml} - d_{sl}(p)]^2 \; ] \qquad (4)$$

where the first component is the sum of weighted squared differences between the measured and simulated (multiphysics, MP, numerical model contribution) time-dependent state variables, the second component is the sum of weighted squared differences between the measured and simulated (surrogate machine learning, ML, model contributions) time-dependent state variables, and weighted squared differences in preferred conditions among adjacent layers (Tikhonov constraint); λ is the Lagrange multiplier that is applied to perform nonlinear parameter estimation as an unconstrained problem (Doherty, 2008); ψ, T, C are pressure head, temperature, and concentration measurements; *d* is the vector of regularization conditions; *p* is the vector of parameter values being estimated (water: $K_{s0x}$; heat: $C_m$, $λ_{slt}$,; solute: $α_L$, $ρ_b$); m and s are subscripts indicating measured and simulated values; NP, NT, and NC are number of pressure head, temperature, and concentration measurements; NR is the number of regularization conditions; $w_{ψi}$, $w_{Ti}$, $w_{Ck}$, are weights for pressure head, temperature, and concentration measurements that are determined iteratively to form a single population with a uniform variance (Gailey and Gorelick, 1991); $w_{rl}$ are regularization weights that are estimated as part of the nonlinear regression process (Doherty, 2008). To enhance the rate of numerical convergence and prevent parameters from becoming negative, all parameters are log transformed.

The Tikhonov regularization (Tikhonov and Arsenin, 1977) information is introduced as preferred conditions (soft prior information) that are defined by equations where the difference between logarithms of selected model layer parameters is initially set to zero (homogeneous condition). The preferred information is considered soft because prior values will change if there is sufficient information content in the calibration-constraint measurements that promote estimation of parameter values with a continued reduction in the overall objective function. In this way, the model structure changes as part of the nonlinear least-squares estimation process with mutual information maximized to reduce the objective function. Upon satisfying a user specified convergence criteria, the nonlinear regression procedure is said to have calibrated the joint multiphysics decision tree model; therefore, the multiphysics numerical is calibrated and surrogate supervised machine learning models considered to be of reduced order. The reader is referred to Doherty (2008) for a review of the Gauss-Levenberg-Marquardt least-squares algorithm used herein.

## 3. RESULTS AND DISCUSSION

### 3.1 Joint multiphysics-decision tree learning parameter estimation

In general (excluding the estimate of fractional sand for layer 1), the parameter estimation results reveal that the joint multiphysics-decision tree learning algorithm outperformed the multiphysics numerical inversion for all properties in each of the model layers (Table 6). For example, the respective average layer 1 (upper gravel layer from 0 m to 3 m depth) percentage differences between actual and estimated multiphysics joint inversion and joint multiphysics-decision tree learning property values are 30.6% and 0.81%, the respective average layer 2 (middle silty layer from 3 m to 5 m depth) differences between actual and estimated multiphysics joint inversion and joint multiphysics-decision tree learning property values are 15.2% and 0.09%, and the respective average layer 3 (bottom sand layer from 5 m to 14m depth) differences between actual and estimated multiphysics joint inversion and joint multiphysics-decision tree learning property values are -3.27% and 0.58%. Given that the multiphysics-informed decision tree learning algorithm resulted in about one order of magnitude less estimation error than the multiphysics numerical inversion suggests that models in which machine-learning methods are introduced into the inversion provide significantly better parameter estimates (for this managed aquifer recharge model, the average prediction error <1% for all layers). Our results agree with others that found improvements in parameter estimates when using a physics-informed deep learning implementation (He et al., 2020; Tartakovsky et al., 2020). Stated another way, regularization of the multiphysics numerical model using machine or deep learning can reduce parameter estimation and prediction uncertainty.

In addition to improving parameter estimation, the multiphysics-decision tree learning algorithm reduced the computational burden associated with Jacobian calculations by about 90% when compared to the multiphysics numerical inversion (Table 7). For example, using the multiphysics joint inversion to estimate 10 properties with 405 observations (1% split of the field measurements) required 113,400 sensitivity calculations over 28 iterations (4050 sensitivity calculations per iteration). In switching to the joint multiphysics- decision tree learning inversion, the total number of calculations is reduced to 12,150 with convergence occurring after only 3 nonlinear iterations. This result suggests that there is the potential for time savings and possibility to move numerical modeling of field sites from the high-performance computing environment to desktop workstations.

## 3.2 Reduced-order model simulations

A natural outcome following convergence of the multiphysics-informed machine learning algorithm is that the calibrated set of surrogate machine learning models comprise a reduced-order model. The reduced-order model reflects the supervised machine-learning models informed by the explicitly-coupled nonlinear multiphysics partial differential equations (1–3). Comparison of the numerical multiphysics pressure head, temperature, and concentration simulations (left panel) with the supervised reduced-order model (pressure head predicted by Random Forest, and temperature and concentration predicted by Ensemble Gradient Boosting) simulations (right panel) reveals excellent correspondence (Figure 8). This visual finding is supported quantitatively by the corresponding Nash-Sutcliffe model efficiency coefficients (Nash and Sutcliffe, 1970) calculated (Table 8) among the multiphysics numerical and reduced-order model simulations for state variables > 0.93 (e.g. pressure head = 0.95, temperature = 0.98, concentration= 0.93). The Nash–Sutcliffe efficiency is calculated as one minus the ratio of the error variance of the modeled time-series divided by the variance of the observed time-series. In the situation of a perfect model with an estimation error variance equal to zero, the resulting Nash-Sutcliffe model efficiency coefficient equals 1. These results suggest that the joint multiphysics-decision tree learning (random forest and ensemble gradient boosting) models can be used as a reduced order model to evaluate changing boundary conditions on the distribution of subsurface state variables at this managed aquifer recharge site without the need of a multiphysics numerical model and high-performance computing. Inspection of the selected time-series profiles reveals the effect of stratigraphic controls on the transport of mass and energy. Specifically, the nonlinear pressure head and temperature changes occur from the streambed surface to the water table. The majority change in concentration occurs from the surface to a depth of about 4 m. That is, the pressure head and heat propagated through all three layers, but the bulk of solute transport was confined to the uppermost layer.

The concept of supervised reduced-order modeling is extended to unsupervised reduced order modeling (Figure 9). For this case, the *unsupervised* machine learning algorithm being evaluated is a modified version of the Self-Organized Map (Kohonen, 2001). Training of this unsupervised learning algorithm is undertaken using the joint multiphysics-decision tree learning property estimates from section 3.1 (Table 6), stochastic boundary conditions, and subsurface state variables. The training (N=31,428) and independent testing (N=7,857) records include state variables (pressure head, temperature, and concentration) and predictor variables (water: saturated hydraulic conductivity; heat: thermal conductivity; solute: bulk density and longitudinal dispersivity; time and depth). The selection of these predictor variables (features) was determined based the identification of influential parameters (those with mutual information about those parameters) using the embedded wrapper learn-heuristics algorithm (Table 4). The training involved competitive learning with a rough phase (initial radius = 144 units, final radius = 36 units, and training length = 20 units) and fine phase (initial radius = 36 units, final radius = 1 units, and training length = 400 units) with initial and learning rates of 0.5 and 0.05 that decay linearly down to $10^{-5}$, and Gaussian neighborhood function that decreased exponentially (decay rate of $10^{-3}$ iteration$^{-1}$). The various network dimensions (rows x columns) evaluated (54 x 48, 108 x 94, 136 x 118, and 162 x 142) yielded quantization errors (average distance between each data vector and its best matching unit vector) on the order of 0.01, and final topographic errors (the proportion of all data vectors for which first (and second) best matching unit vectors are not adjacent units) on the order of 0.1. The inability to improve predictions with increasing network dimensions implies that the smallest size network as determined using the heuristic formula (Kohonen, 2001) can be considered optimal for independent testing.

The presentation of independent test sets (N=7,939 records; 20% splits from original stochastic data set) to the trained unsupervised learning model results in the simultaneous prediction of pressure head, temperature, and concentration (Figure 9). Inspecting these results reveals unbiased predictions with near-zero mean residuals (e.g. mean pressure head residual = -0.006 m, mean temperature residual = 0.09 $^{0}$C, mean concentration residual = -0.006 mol kg$^{-1}$). These metrics indicate that the Self-Organized Map is an excellent predictor of subsurface state variables when using the multiphysics-decision tree learning layer property estimates and stochastic boundary conditions to train the model. The visual magnitude of nonlinear prediction uncertainty after 1 day are computed following independent testing with 30 random split sets resulted in the simultaneous predictions of pressure head, temperature, and concentration from which 25$^{th}$, 50$^{th}$, and 75$^{th}$ percentiles are determined and presented in Figure 10. These subsurface profiles of state variables compare favorably with the multiphysics numerical and supervised reduced order model simulations after 1 day shown in Figure 8. These findings suggest that while the multiphysics-decision tree learning estimates are likely to be initiated using a high-performance computing environment, the unsupervised reduced order model need only be used on a personal desktop computer to evaluate changing boundary conditions and subsurface properties on the temporal distribution of pressure head, temperature, and concentration in the managed aquifer recharge basin.

## 4. CONCLUSIONS

In this study, the proposed multiphysics-decision tree learning algorithm is successfully developed and applied to the (conceptual) variably saturated recharge basin in southern California. Major strengths of this algorithm are the improved computational efficiency (e.g., the reduced number Jacobian sensitivity calculations and iterations until convergence), improved accuracy in the parameter estimation of water, heat, and solute transport parameters (relative to the multiphysics joint inversion), and comparable reduced order model simulations (relative to the calibrated multiphysics model) of spatiotemporal pressure head, temperature, and concentration. The supervised reduced order model honoring the multiphysics equations can be represented by a set of machine learning algorithms for coupled predictions of state variables, such as the Random Forest regressor for pressure head and the ensemble gradient boosting for temperature and concentration. The unsupervised reduced order model that honors multiphysics by training the algorithm with the multiphysics-decision tree learning estimated water, heat, and solute transport properties (features). The selection of training properties is determined based the identification of influential parameters (those with mutual information about those parameters) by feature selection and learn-heuristics. Benefits of this reduced order model are that changes to infiltrating water, heat, and solute (surface boundary conditions) can be efficiently studied for their effects on the redistribution and uncertainty of pressure head, temperature, and concentration (assuming the same or different subsurface properties). The results presented herein provide impetus for the extension of the physics-informed machine learning to the estimation of subsurface geophysical properties and joint geophysical, and joint hydrogeologic-geophysical properties.

**ACKNOWLEDGEMENTS**

The research was funded by the Laboratory Directed Research and Development Program at Pacific Northwest National Laboratory (PNNL) (PNNL contract NF2564). The authors also wish to thank George Muntean and Zhangshuan (Jason) Hou of the PNNL for their guidance and support during this research study.

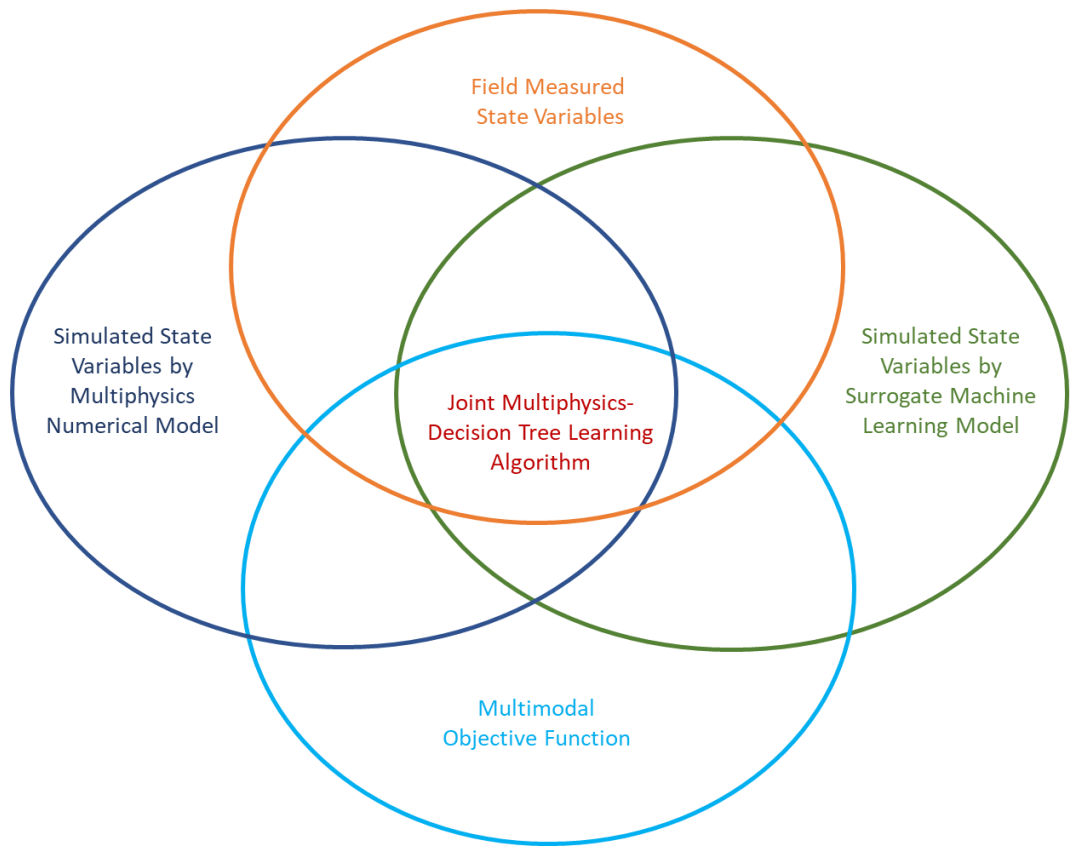

Figure 1. General components of the multiphysics–decision tree learning algorithm for variably saturated subsurface transport: field measured state variables, simulated state variables by multiphysics numerical model, simulated state variables by surrogate machine learning models, multimodal objective function (field measured and simulated state variables). State variables include pressure head, temperature, and solute concentration.

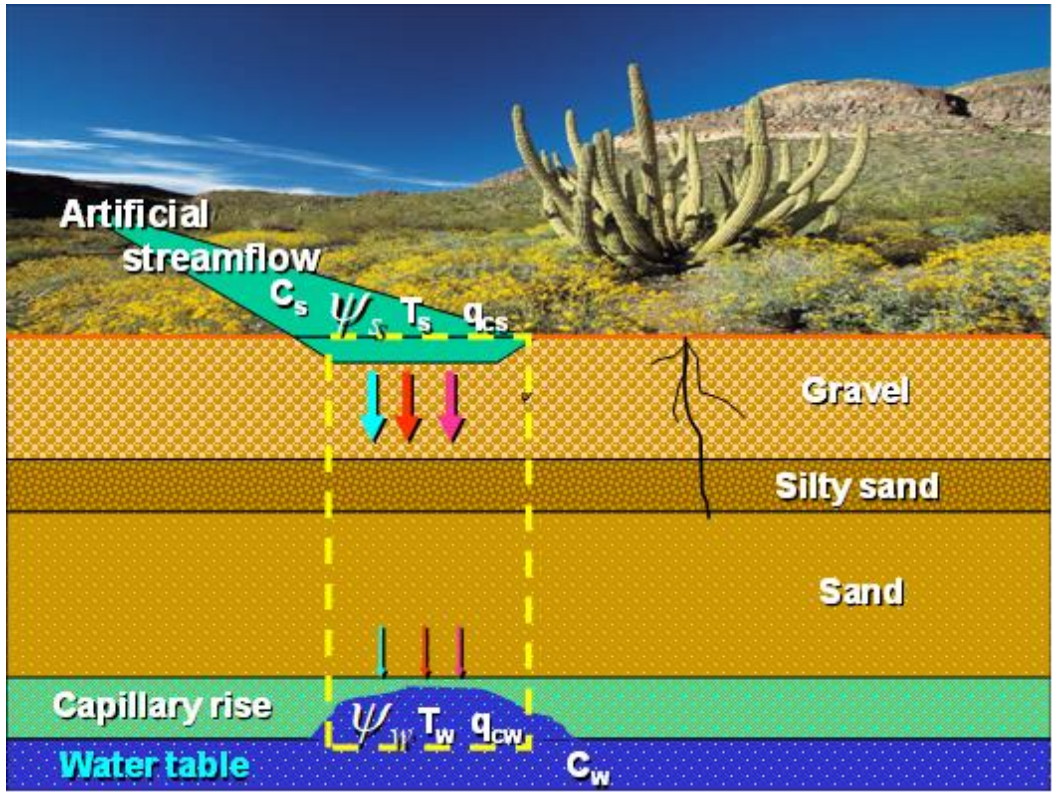

Figure 2. Schematic of the managed aquifer recharge with boundary conditions. $C_s$ = far field concentration of stream, $\psi_s$ = pressure head of stream, $T_s$ = temperature of stream, $q_{cs}$ = convective solute flux at stream , $\psi_w$ , $T_w$ = temperature at water table, $q_{cw}$ = convective solute flux at water table, $C_w$ = far field concentration of water table.

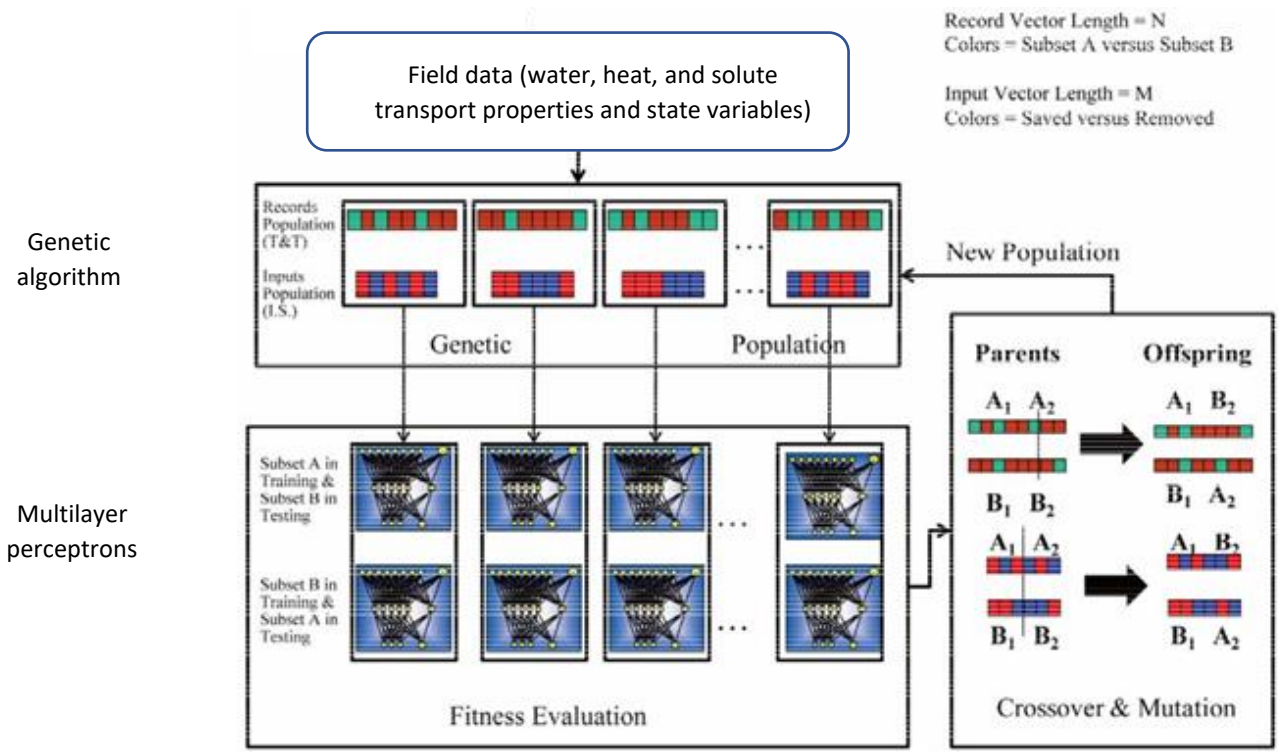

Figure 3. The Training Wih Input Selction and Testing (TWIST) learn heuristics algorithm used for feature selection (after Buscema et al., 2013).

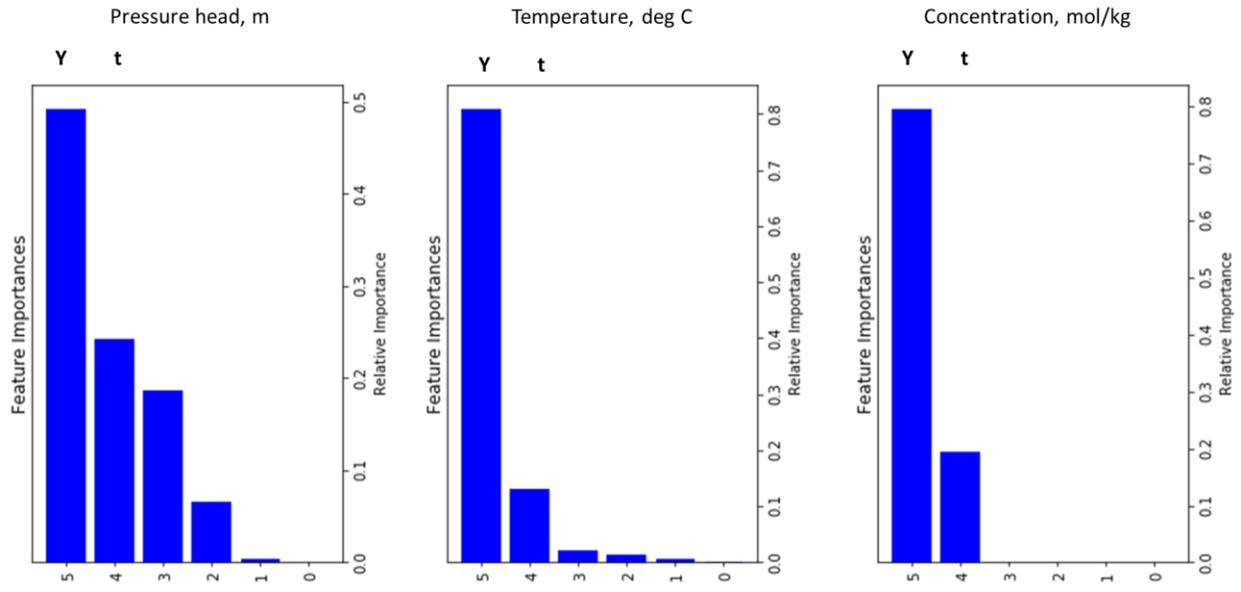

Figure 4. Feature importance for three trained Random Forest models for pressure head, temperature, and concentration reveals sensitivity to time and distance and **insensitivity to the four physical properties,** where Y = depth (m), and t = time (d).

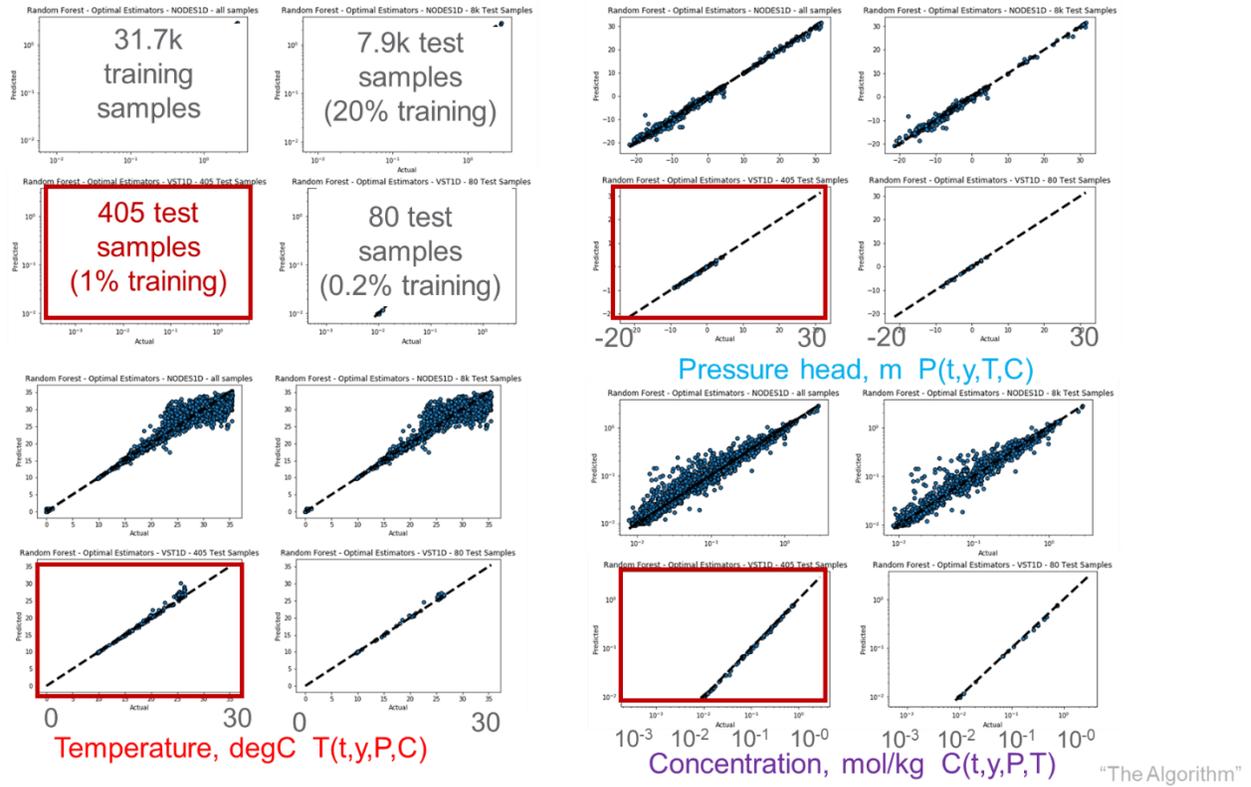

Figure 5. Independent Random Forest model predictions of pressure head (m), temperature ($^0$C), and concentration (mol kg$^{-1}$). Upper left quadrant: panels indicate number of samples used in training or testing; N=31,751 training samples, N=7,939 test samples (20% original data), 405 test samples (1% original data), 80 test samples (0.2% original data); upper right quadrant: P = pressure head (training and testing); lower left quadrant: T = temperature, lower right quadrant: C = concentration, where t = time (d), and y = depth (m).

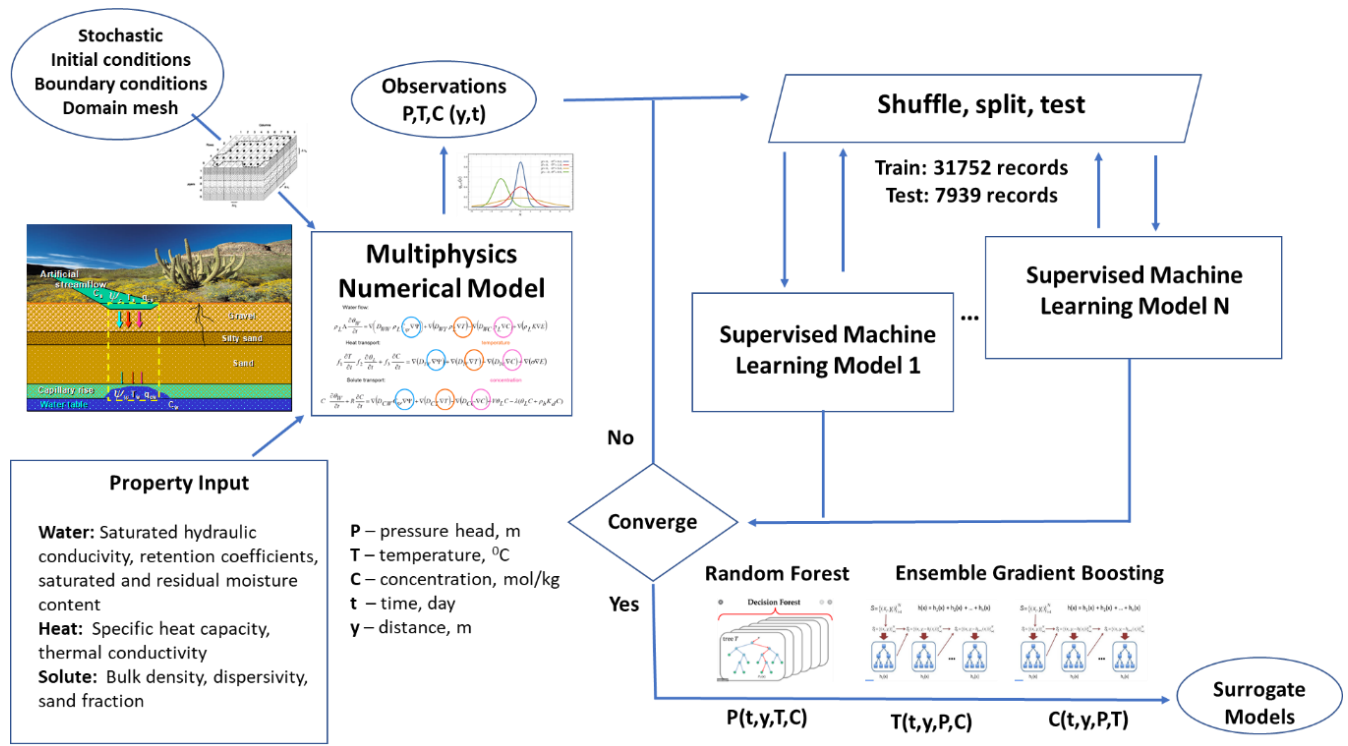

Figure 6. The workflow used to identify suitable surrogate models for predicting pressure head, temperature and concentration as part of the joint multiphysics-decision tree learning algorithm.

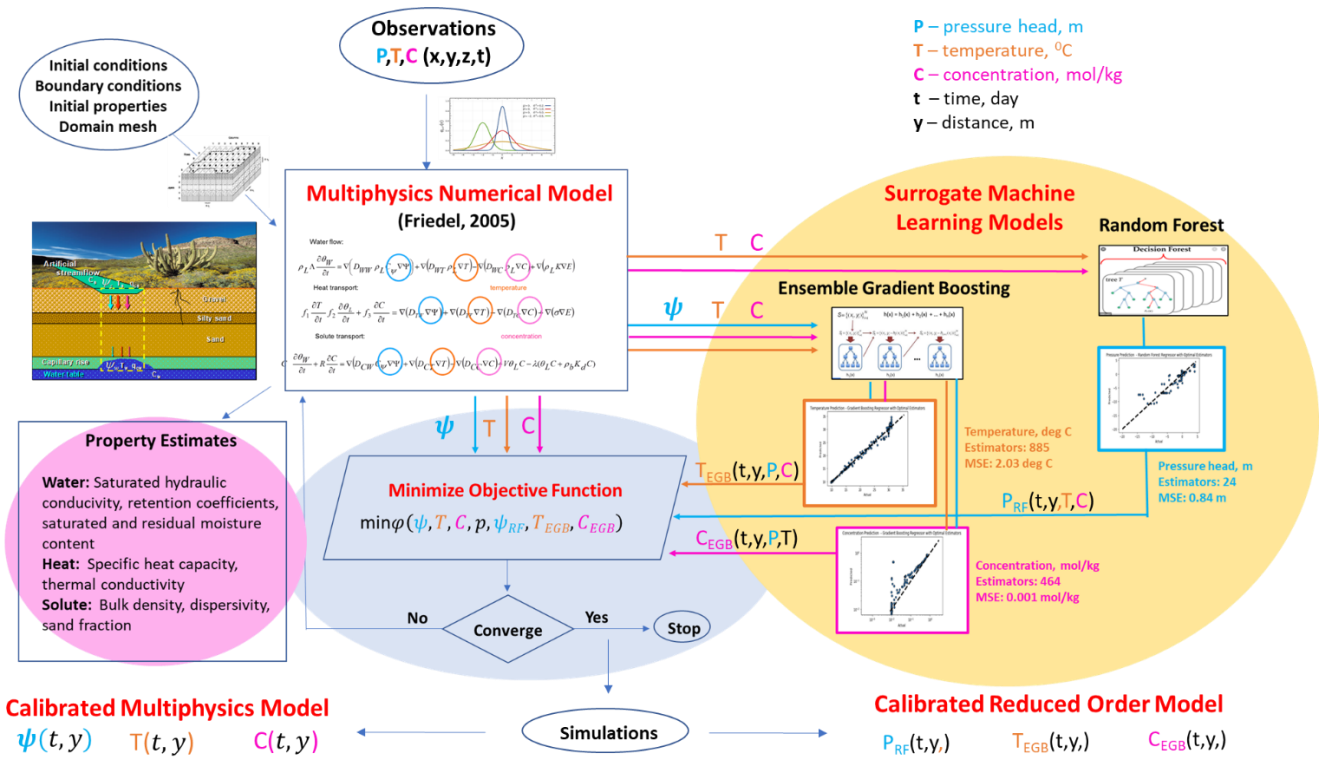

Figure 7. The multiphysics-decision tree learning workflow used to estimate water, heat, and solute transport properties and provide simulation using the calibrated multiphysics model and calibrated

supervised reduced order model (physics informed supervised machine learning models).

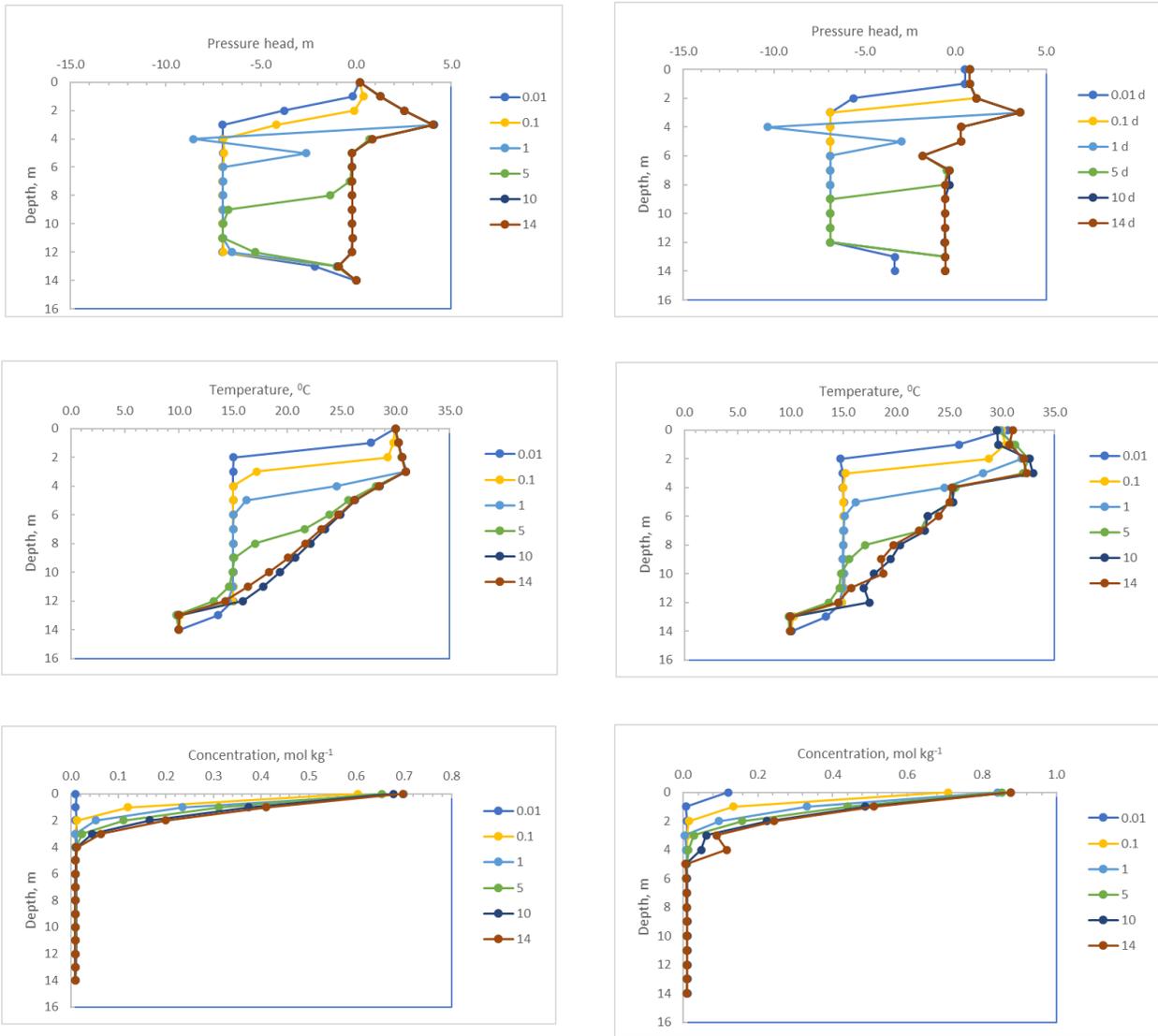

Figure 8. Comparison of simulated pressure head, temperature and concentration using the calibrated numerical multiphysics model (left) and calibrated supervised reduced-order model (right). The supervised reduced-order model predictions are undertaken by trained Random Forest (pressure head) as a function of time, depth, temperature, and concentration; Ensemble Gradient Boosting (temperature) as a function of time, depth, pressure head and concentration; and Ensemble Gradient Boosting (concentration) as a function of time, depth, pressure head and temperature.

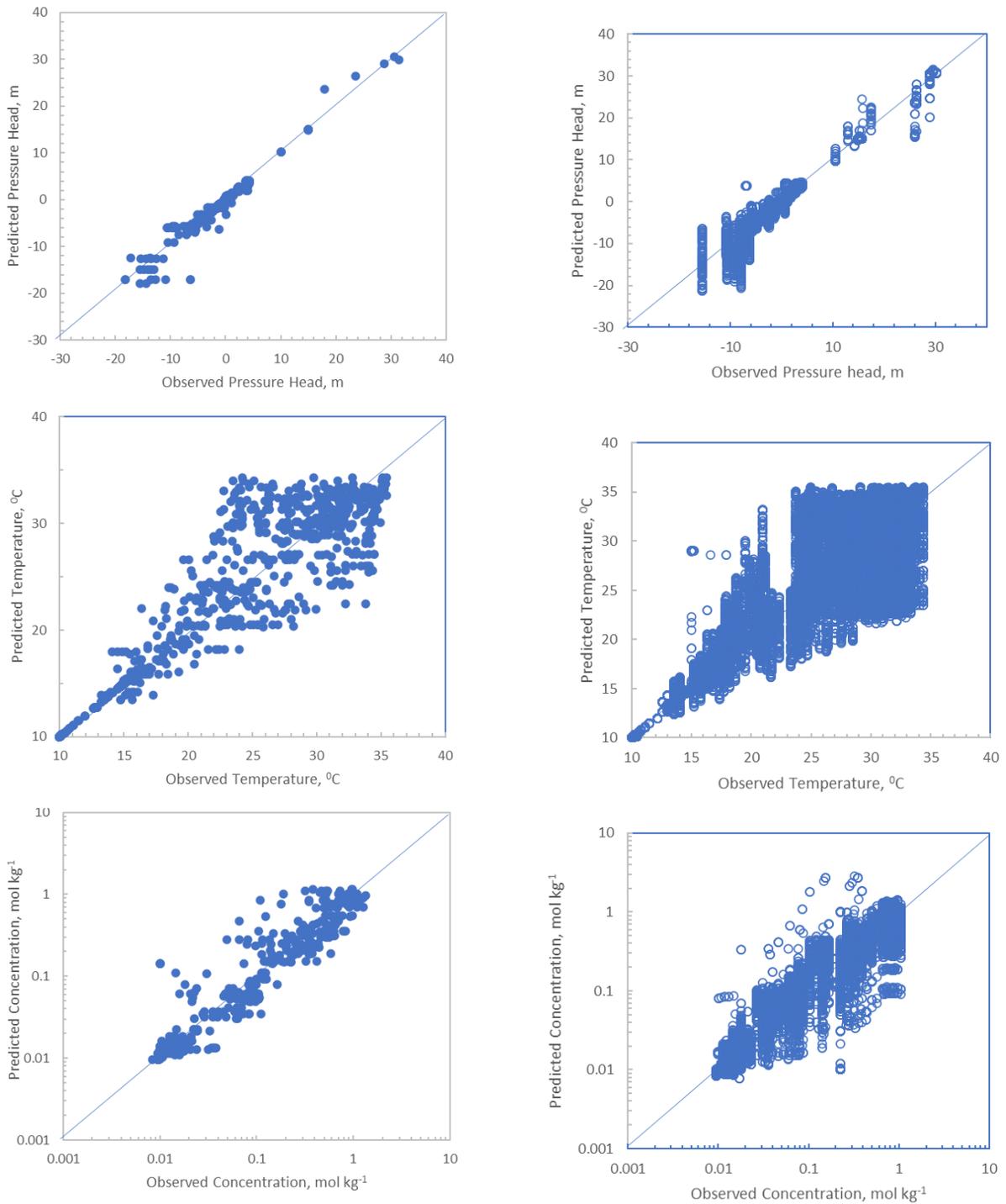

Figure 9. Comparison of reduce-order model predictions to observations for multiphysics state variables. The reduced-order model is an unsupervised machine learning algorithm (self-organizing map) trained with state variables and properties estimated using the multiphysics-decision tree learning algorithm. The training (N=31,751 records) and independent testing (random sets each with N=7,939 records) data included state (pressure head, temperature, and concentration) and predictor (water: saturated hydraulic

conductivity; heat: thermal conductivity; solute: bulk density and longitudinal dispersivity variables. Left panel presents a random testing set; right panel presents 30 random testing sets.

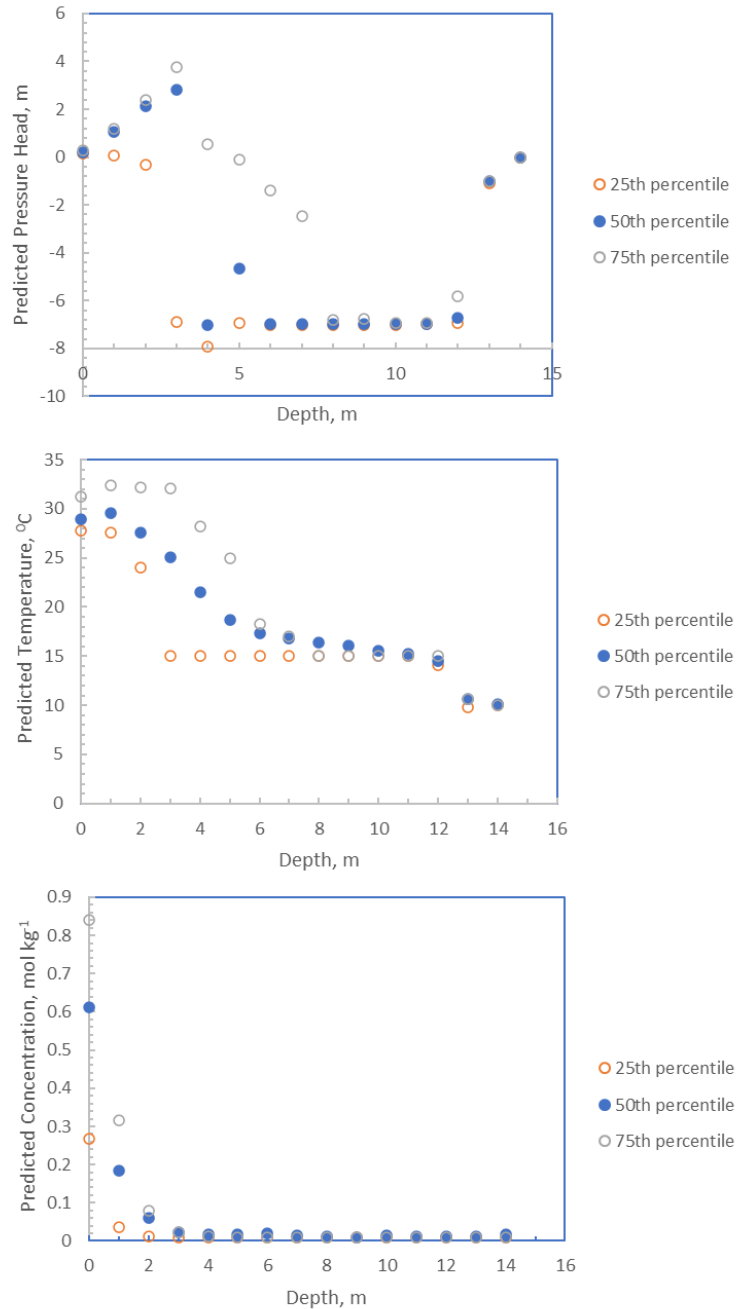

Figure 10. Predicted variation in pressure head, temperature, and concentration after 1 day using the unsupervised reduced-order model and stochastic surface boundary conditions. The reduced-order model is an unsupervised machine learning algorithm (self-organizing map) trained (N=31,751 records; 80% split from original stochastic data set) with state variables and properties estimated using the joint multiphysics-decision tree learning algorithm. The independent testing with 30 random subsets (N=7,939

records per subset; 20% splits from original stochastic data set) produced simultaneous predictions of pressure head, temperature, and concentration from which $25^{th}$, $50^{th}$, and $75^{th}$ percentiles are shown. The unbiased subsurface variation in state variable predictions reflects stochastic nature of the streambed boundary conditions at the conceptual managed aquifer recharge basin.

Table 1. Summary of subsurface water, heat, and solute transport properties at the managed aquifer recharge basin. Water: kxs = saturated hydraulic conductivity, alpha = scaling parameter of the Van Genuchten retention model, n = shape parameter of the Van Genuchten retention model, tr = residual water content, ts = saturated water content - porosity; Heat: cps = volumetric heat capacity, lambdas = thermal conductivity, Solute: rhob = bulk density, al = longitudinal dispersivity, xsnd = fractional sand content.

| Layer | Y (m) | kxs (m day$^{-1}$) | alpha (m$^{-1}$) | n | tr (m$^3$ m$^{-3}$) | ts (m$^3$ m$^{-3}$) | cps (cal m$^{-3}$ $^0$C$^{-1}$) | lambdas (cal m$^{-1}$ $^0$C$^{-1}$ d$^{-1}$) | rhob (kg m-3) | al (m) | xsnd (m$^3$ m$^{-3}$) |
|---|---|---|---|---|---|---|---|---|---|---|---|
| 1 | 0 to 3 | 4.8 | 13.49 | 3.13 | 0.03 | 0.6 | 0.48 | 12 | 1.7 | 0.4 | 0.9 |
| 2 | 3 to 5 | 0.07 | 0.2 | 1.73 | 0.09 | 0.3 | 0.48 | 12 | 1.45 | 0.2 | 0.8 |
| 3 | 5 to 14 | 2.4 | 6.37 | 1.65 | 0.13 | 0.3 | 0.48 | 12 | 1.65 | 0.9 | 0.6 |

Table 2. Stochastic boundary conditions simulated for application to stream boundary in managed aquifer recharge basin.

|  | Original | | | Down selected | | |
|---|---|---|---|---|---|---|
| State Variable | Pressure Head (m) | Temperature (⁰C) | Concentration (mol/kg) | Pressure Head (m) | Temperature (⁰C) | Concentration (mol/kg) |
| Trials | 1,000 | 1,000 | 1,000 | 100 | 100 | 100 |
| Base Case | 0.2 | 30 | 1 | 0.2 | 30 | 1 |
| Mean | 0.19 | 30.1 | 1.01 | 0.22 | 29.8 | 0.98 |
| Standard Deviation | 0.25 | 3.1 | 0.32 | 0.23 | 2.9 | 0.29 |
| Minimum | -0.57 | 20.3 | 0.01 | -0.52 | 23.0 | 0.30 |
| 25% | 0.03 | 28.1 | 0.81 | 0.06 | 27.8 | 0.77 |
| 50% | 0.19 | 30.0 | 1.00 | 0.27 | 29.9 | 0.99 |
| 75% | 0.35 | 32.1 | 1.21 | 0.38 | 31.8 | 1.2 |
| Maximum | 0.97 | 39.8 | 1.99 | 0.97 | 39.1 | 1.9 |

Table 3. Simulated state variables following application of stochastic conditions to stream boundary in managed aquifer recharge basin (*N=39,285k*): Random shuffle and split: training *(N=31,428)* and testing *(N=7,857*) and (*N=405*).

| Statistic | Training | | | Testing | | |
|---|---|---|---|---|---|---|
| | Pressure Head (m) | Temperature ($^0$C) | Concentration (mol kg$^{-1}$) | Pressure Head (m) | Temperature ($^0$C) | Concentration (mol kg$^{-1}$) |
| Trials | 31,428 | 31,428 | 31,428 | 7,857 | 7,857 | 7,857 |
| Base Case | 0.2 | 30 | 1 | 0.2 | 30 | 1 |
| Mean | -3.09 | 18.67 | 0.070 | -3.20 | 18.60 | 0.070 |
| Standard Deviation | 4.61 | 7.52 | 0.185 | 3.90 | 7.50 | 0.190 |
| Minimum | -22.0 | 0.01 | 0.008 | -20.0 | 0.01 | 0.008 |
| 25% | -7.00 | 15.00 | 0.010 | -6.99 | 14.99 | 0.009 |
| 50% | -2.82 | 15.0 | 0.010 | -3.46 | 15.0 | 0.010 |
| 75% | 0.00 | 24.4 | 0.010 | 0.00 | 24.2 | 0.010 |
| Maximum | 4.6 | 35.5 | 2.88 | 4.5 | 35.5 | 2.83 |

| Statistic | Testing | | | Testing | | |
|---|---|---|---|---|---|---|
| | Pressure Head (m) | Temperature ($^0$C) | Concentration (mol kg$^{-1}$) | Pressure Head (m) | Temperature ($^0$C) | Concentration (mol kg$^{-1}$) |
| Trials | 7,857 | 7,857 | 7,857 | 405 | 405 | 405 |
| Base Case | 0.2 | 30 | 1 | 0.2 | 30 | 1 |
| Mean | -3.20 | 18.60 | 0.070 | -3.31 | 18.45 | 0.062 |
| Standard Deviation | 3.90 | 7.50 | 0.190 | 3.98 | 7.29 | 0.186 |
| Minimum | -20.0 | 0.01 | 0.008 | -14.5 | 9.77 | 0.009 |
| 25% | -6.99 | 14.99 | 0.009 | -6.99 | 14.99 | 0.009 |
| 50% | -3.46 | 15.0 | 0.010 | -3.83 | 15.0 | 0.010 |
| 75% | 0.00 | 24.2 | 0.010 | 0.00 | 23.3 | 0.010 |
| Maximum | 4.5 | 35.5 | 2.83 | 4.1 | 35.0 | 1.09 |

Table 4. Feature selection using the TWIST embedded wrapper algorithm (Buscema et al., 2013) to identify robust model parameters based on a modal threshold of 0.67.

| Number | Parameter type | Predictor variable | Count N | Fraction | Threshold >0.67 | Description |
|---|---|---|---|---|---|---|
| 1 | All | TIME | 15 | 1 | X | Time (d) |
| 2 | All | Y | 15 | 1 | X | Vertical distance (m) |
| 3 | Water | kxs | 15 | 1 | X | Hydraulic conductivity (m $d^{-1}$) |
| 4 | Water | alpha | 3 | 0.2 | - | Scaling parameter of the Van Genuchten retention model ($m^{-1}$) |
| 5 | Water | n | 7 | 0.47 | - | Shape parameter of the Van Genuchten retention model (dimensionless) |
| 6 | Water | tr | 9 | 0.6 | - | Residual water content ($m^3 \, m^{-3}$) |
| 7 | Water | ts | 2 | 0.13 | - | Saturated water content - porosity ($m^3 \, m^{-3}$) |
| 8 | Heat | cps | 8 | 0.53 | - | Volumetric heat capacity (cal $m^{-3} \, ^0C^{-1}$) |
| 9 | Heat | lambdas | 11 | 0.73 | X | Thermal conductivity (cal $m^{-1} \, ^0C^{-1} \, d^{-1}$) |
| 10 | Solute | rhob | 10 | 0.67 | X | Bulk density (kg $m^{-3}$) |
| 11 | Solute | al | 11 | 0.73 | X | Longitudinal dispersivity (m) |
| 12 | Solute | xsnd | 2 | 0.13 | - | Fractional sand content ($m^3 \, m^{-3}$) |

X indicates that mode exceeds the threshold (>0.67) reflecting feature modes are within 1 standard deviation of the mean.

Table 5. Random Forest regressor feature importance for pressure head, temperature, and concentration models.

| Predictor variable | Dependent variable | | |
|---|---|---|---|
| | Pressure head (m) | Temperature ($^0$C) | Concentration (mol kg$^{-1}$) |
| Time (d) | 0.11 | 0.00 | 0.14 |
| Depth (m) | 0.02 | 0.13 | 0.81 |
| Pressure head (m) | NA | 0.07 | 0.20 |
| Temperature ($^0$C) | 0.73 | NA | 0.02 |
| Concentration (mol kg$^{-1}$) | 0.24 | 0.79 | NA |

Table 6. Comparison of estimated layer properties using multiphysics inversion and the multiphysics--decision tree learning inversion algorithms.

| Layer | Lithology | Property type | Property variable | Actual value | Multiphysics inversion Estimated value (Friedel, 2005) | Joint multiphysics-decision tree learning inversion Estimated value | Joint multiphysics-decision tree learning inversion lower limit | Joint multiphysics-decision tree learning inversion upper limit | Multiphysics inversion Difference from actual, % | Joint multiphysics-decision tree learning inversion Difference from actual, % | Property name |
|---|---|---|---|---|---|---|---|---|---|---|---|
| 1 | Gravel | Water | kxs1 | 4.80 | 5.239 | 4.815 | 4.237 | 5.472 | 9.1 | 0.3 | Hydraulic conductivity (m d$^{-1}$) |
| 1 | Gravel | Water | alpha1 | 13.49 | 12.566 | 13.404 | 10.261 | 17.510 | -6.8 | -0.6 | Scaling parameter for Van Genuchten retention model (m$^{-1}$) |
| 1 | Gravel | Water | n1 | 3.13 | 3.171 | 3.118 | 2.408 | 4.037 | 1.3 | -0.4 | Shape parameter of the Van Genuchten retention model (dimensionless) |
| 1 | Gravel | Water | tr1 | 0.03 | 0.075 | 0.027 | 0.020 | 0.036 | 188.5 | 2.5 | Residual water content (m$^3$ m$^{-3}$) |
| 1 | Gravel | Water | ts1 | 0.60 | 0.698 | 0.599 | 0.503 | 0.715 | 16.3 | -0.1 | Saturated water content (m$^3$ m$^{-3}$) |
| 1 | Gravel | Heat | cps1 | 0.48 | 0.585 | 0.505 | 0.246 | 1.037 | 21.9 | 5.3 | Volumetric heat capacity (cal m$^{-3}$ $^0$C$^{-1}$) |
| 1 | Gravel | Heat | lambdas1 | 12.0 | 14.98 | 11.97 | 3.042 | 47.08 | 24.9 | -0.3 | Thermal conductivity (cal m$^{-1}$ $^0$C$^{-1}$ d$^{-1}$) |
| 1 | Gravel | Solute | rhob1 | 1.70 | 1.820 | 1.705 | 1.337 | 2.173 | 7.1 | 0.3 | Bulk density (kg m$^{-3}$) |
| 1 | Gravel | Solute | al1 | 0.40 | 0.454 | 0.402 | 0.287 | 0.561 | 13.5 | 0.4 | Longitudinal dispersivity (m) |
| 1 | Gravel | Solute | xsnd1 | 0.90 | 0.432 | 0.050 | 0.023 | 0.106 | -52.0 | -94.5 | Fractional sand content (m$^3$ m$^{-3}$) |
| 2 | Silty-sand | Water | kxs2 | 0.07 | 0.084 | 0.072 | 0.061 | 0.086 | 16.7 | 0.5 | Hydraulic conductivity (m d$^{-1}$) |
| 2 | Silty-sand | Water | alpha2 | 0.20 | 0.223 | 0.199 | 0.139 | 0.285 | 11.5 | -0.4 | Scaling parameter for Van Genuchten retention model (m$^{-1}$) |
| 2 | Silty-sand | Water | n2 | 1.73 | 1.794 | 1.740 | 1.387 | 2.181 | 3.7 | 0.6 | Shape parameter of the Van Genuchten retention model (dimensionless) |
| 2 | Silty-sand | Water | tr2 | 0.09 | 0.099 | 0.088 | 0.050 | 0.158 | 12.5 | 0.5 | Residual water content (m$^3$ m$^{-3}$) |
| 2 | Silty-sand | Water | ts2 | 0.30 | 0.333 | 0.302 | 0.217 | 0.419 | 11.0 | 0.5 | Saturated water content (m$^3$ m$^{-3}$) |
| 2 | Silty-sand | Heat | cps2 | 0.48 | 0.604 | 0.480 | 0.260 | 0.886 | 25.8 | 0.0 | Volumetric heat capacity (cal m$^{-3}$ $^0$C$^{-1}$) |
| 2 | Silty-sand | Heat | lambdas2 | 12.0 | 14.60 | 11.90 | 5.614 | 25.24 | 21.6 | -0.8 | Thermal conductivity (cal m$^{-1}$ $^0$C$^{-1}$ d$^{-1}$) |
| 2 | Silty-sand | Solute | rhob2 | 1.45 | 1.389 | 1.456 | 1.046 | 2.026 | -4.2 | 0.4 | Bulk density (kg m$^{-3}$) |
| 2 | Silty-sand | Solute | al2 | 0.20 | 0.303 | 0.200 | 0.042 | 0.938 | 51.5 | -0.2 | Longitudinal dispersivity (m) |
| 2 | Silty-sand | Solute | xslt2 | 0.80 | 0.817 | 0.799 | 0.308 | 2.069 | 2.1 | -0.2 | Fractional sand content (m$^3$ m$^{-3}$) |
| 3 | Sand | Water | kxs3 | 2.40 | 2.604 | 2.441 | 1.778 | 3.351 | 8.5 | 1.7 | Hydraulic conductivity (m d$^{-1}$) |
| 3 | Sand | Water | alpha3 | 6.37 | 6.215 | 6.399 | 4.972 | 8.235 | -2.4 | 0.5 | Scaling parameter for Van Genuchten retention model (m$^{-1}$) |
| 3 | Sand | Water | n3 | 1.65 | 1.718 | 1.652 | 1.218 | 2.240 | 4.1 | 0.1 | Shape parameter of the Van Genuchten retention model (dimensionless) |
| 3 | Sand | Water | tr3 | 0.13 | 0.125 | 0.130 | 0.091 | 0.186 | -3.8 | 0.2 | Residual water content (m$^3$ m$^{-3}$) |
| 3 | Sand | Water | ts3 | 0.30 | 0.325 | 0.301 | 0.235 | 0.387 | 8.3 | 0.4 | Saturated water content (m$^3$ m$^{-3}$) |
| 3 | Sand | Heat | cps3 | 0.48 | 0.613 | 0.484 | 0.294 | 0.798 | 27.7 | 0.9 | Volumetric heat capacity (cal m$^{-3}$ $^0$C$^{-1}$) |
| 3 | Sand | Heat | lambdas3 | 12.0 | 15.00 | 12.02 | 4.355 | 33.19 | 25.0 | 0.2 | Thermal conductivity (cal m$^{-1}$ $^0$C$^{-1}$ d$^{-1}$) |
| 3 | Sand | Solute | rhob3 | 1.65 | 0.900 | 1.651 | 0.809 | 3.369 | -45.5 | 0.1 | Bulk density (kg m$^{-3}$) |
| 3 | Sand | Solute | al3 | 0.90 | 0.243 | 0.906 | 0.378 | 2.169 | -73.0 | 0.7 | Longitudinal dispersivity (m) |
| 3 | Sand | Solute | xslt3 | 0.60 | 0.710 | 0.606 | 0.299 | 1.227 | 18.3 | 1.0 | Fractional sand content (m$^3$ m$^{-3}$) |

Table 7. Comparison of computational burden for the multiphysics inversion and joint multiphysics-decision tree learning algorithms.

| Method | Feature selection | Training | Observations | Parameters | Derivatives | Iterations | Calculations | Reduction | Case |
|---|---|---|---|---|---|---|---|---|---|
| Multiphysics inversion | No | 1% | 405 | 10 | 4050 | 28 | 113400 | 0 | Base |
| Multiphysics inversion | Yes | 1% | 405 | 10 | 4050 | 21 | 85050 | 0.25 | 2020 |
| Multiphysics informed machine learning | Yes | 1% | 405 | 10 | 4050 | 3 | 12150 | 0.89 | 2020 |

Table 8. Summary of Nash-Sutcliffe model efficiency among the multiphysics numerical and reduced order model (joint multiphysics-decision learning).

| State variable | Number of samples | Nash-Sutcliffe efficiency |
|---|---|---|
| Pressure (m) | 405 | 0.95 |
| Temperature ($^0$C) | 405 | 0.98 |
| Concentration (mol kg$^{-1}$) | 405 | 0.93 |

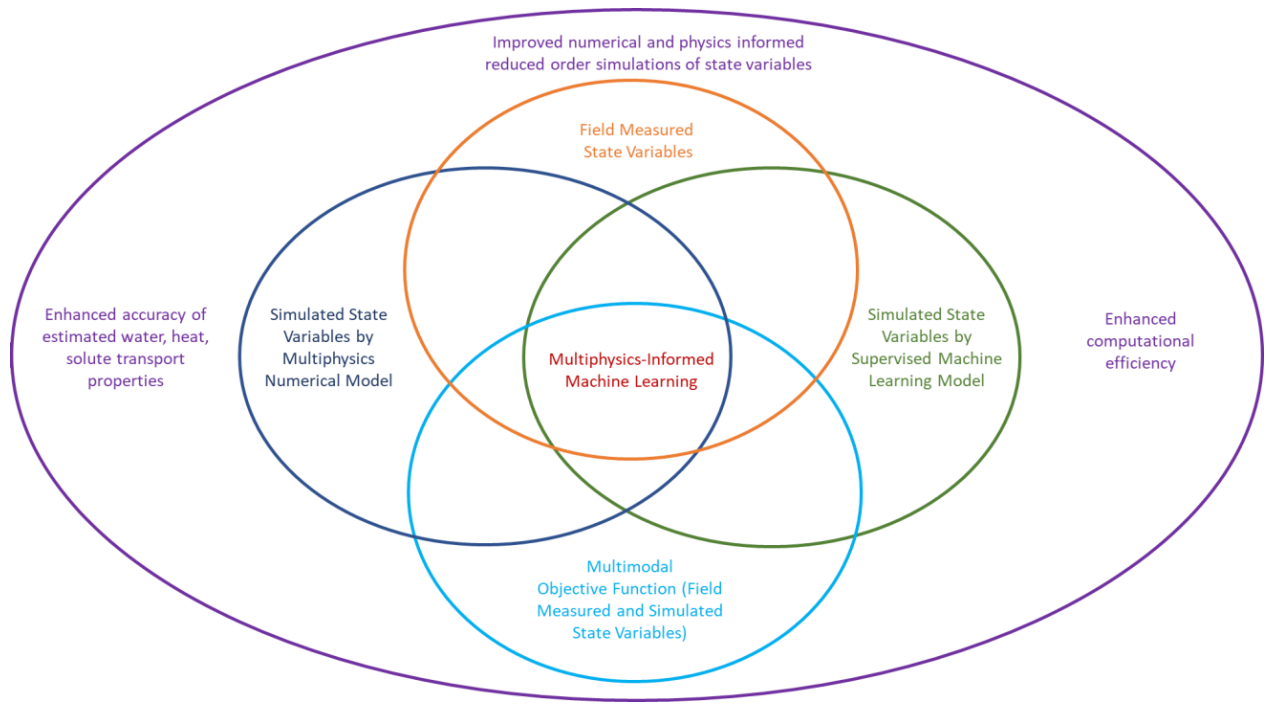

Graphical Abstract: Joint multiphysics-decision tree learning for enhanced accuracy of estimated water, heat, and solute transport properties and improved numerical and joint multiphysics-decision tree learning informed reduced order simulation of state variables (pressure head, temperature, and concentration).